\documentclass[smallextended]{svjour3}       
\smartqed  
\usepackage{graphicx}
\usepackage{amsmath}
\usepackage{url}

\begin{document}

\title{Dynamic Bayesian Games for Adversarial and Defensive Cyber Deception
\thanks{This is a preliminary version of the paper that will appear in the following edited book as a book chapter: 
L. Huang and Q. Zhu, “Deception and Counter-deception Bayesian Game: Adaptive Defense Strategies Against Advanced Persistent Threats for Cyber-physical Systems,” Cyber Deception, E. Al-Shaer, K. Hamlen, J. Wei, and C. Wang (Eds.), Springer, 2018, to appear.}
}


\author{Linan Huang         \and
        Quanyan Zhu 
}


\institute{Linan Huang \at
              Department of Electrical and Computer Engineering, New York University 2 MetroTech Center, Brooklyn, NY, 11201, USA \\
              Tel.: 347-204-2406\\
              \email{lh2328@nyu.edu}           
           \and
          Quanyan Zhu \at
              Department of Electrical and Computer Engineering, New York University 2 MetroTech Center, Brooklyn, NY, 11201, USA \\
              Tel.: 646-997-3371\\
              \email{qz494@nyu.edu}  
}

\date{}
\maketitle
 \begin{abstract}
Security challenges accompany the efficiency. The pervasive integration of information and communications technologies (ICTs) makes cyber-physical systems vulnerable to targeted attacks that are deceptive, persistent, adaptive and strategic. 
Attack instances such as Stuxnet, Dyn, and WannaCry ransomware have shown the insufficiency of  off-the-shelf defensive methods including the firewall and intrusion detection systems. 
Hence, it is essential to design up-to-date security mechanisms that can mitigate the risks despite the successful infiltration and the strategic response of sophisticated attackers. 
In this chapter, we use game theory to model competitive interactions between defenders and attackers. First, we use the static Bayesian game to capture the stealthy and deceptive characteristics of the attacker. 
A random variable called the \textit{type} characterizes users' essences and objectives, e.g., a legitimate user or an attacker. The realization of the user's type is private information due to the cyber deception.
Then, we extend the one-shot simultaneous interaction into the one-shot interaction with asymmetric information structure, i.e., the signaling game. 
Finally, we investigate the multi-stage transition under a case study of Advanced Persistent Threats (APTs) and Tennessee Eastman (TE) process. Two-Sided incomplete information is introduced because the defender can adopt defensive deception techniques such as honey files and honeypots to create sufficient amount of uncertainties for the attacker. 
Throughout this chapter, the analysis of the Nash equilibrium (NE), Bayesian  Nash equilibrium (BNE), and perfect Bayesian Nash equilibrium (PBNE) enables the policy prediction of the adversary and the design of proactive and strategic defenses to deter attackers and mitigate losses.
 
\keywords{Bayesian games \and Multistage transitions \and Advanced Persistent Threats (APTs) \and Cyber deception \and Proactive and strategic defense }
\end{abstract}

\section{Introduction}
\begin{figure}[h]
  \centering
  \includegraphics[width=1\columnwidth]{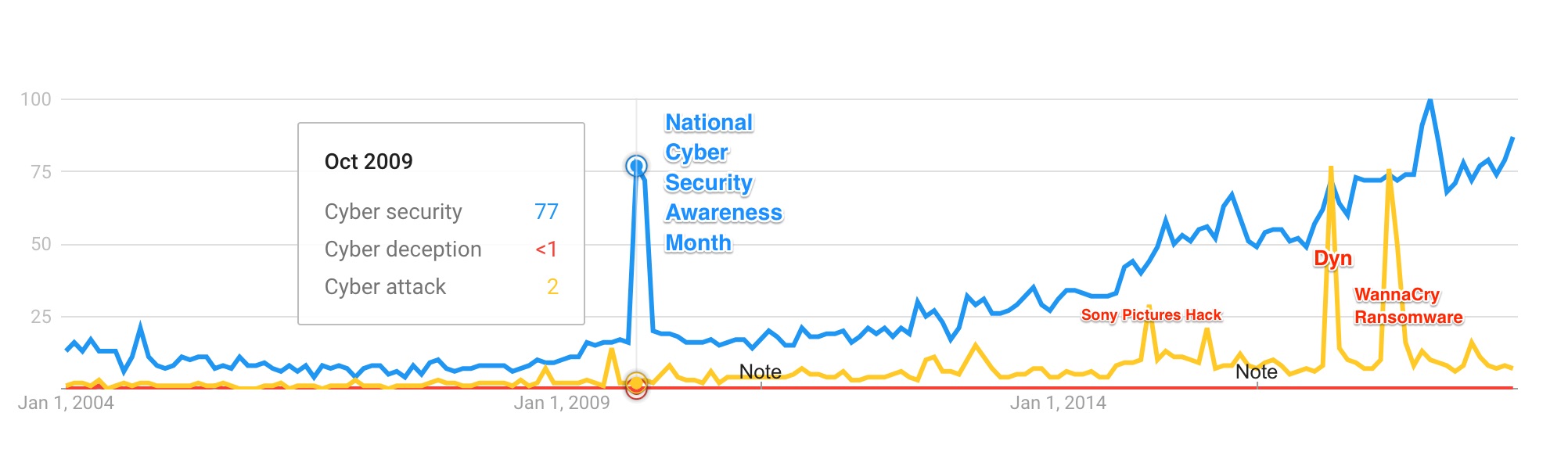}
  \caption{\label{fig: cybertimeline}The search results of three keywords, i.e., the cybersecurity (in blue), the cyber deception (in red) and the cyber attack (in yellow) in the United States from Jan. 2004, to Aug. 2018 via the {\texttt GoogleTrends}. Compared with the blue and yellow line, the cyber deception which endows attackers an information advantage over the defender requires more investigations.  
  Numbers on the $y$-axis represent the search frequency normalized with respect to the highest point on the chart for the given region and time. A value of 100 is the peak popularity.
  }
\end{figure}

The operation of the modern society intensively relies on the Internet services and information and communications technologies (ICTs). Cybersecurity has been an increasing concern as a result of the pervasive integration of ICTs as witnessed in Fig. \ref{fig: cybertimeline}. Every peak of the yellow line corresponds to a cyber attack\footnote{https://en.wikipedia.org/wiki/List\_of\_cyberattacks} and both the frequency and the magnitude which represents the scope of influence has increased, especially in recent years. For example, the Domain Name System (DNS) provider Dyn has become the targeted victim of the multiple distributed denial-of-service (DDoS) attacks in October 2016. The Mirai malware has turned a large number of IoT devices such as printers and IP cameras to bots and causes an estimate of $1.2$ Tbps network flow. More recently in May 2017, the WannaCry ransomware has attacked more than 200,000 computers across 150 countries, with total damages up to billions of dollars. 

One way to contend with the cyber attacks is for the defenders to set up firewalls with pre-defined rules to prevent their internal network from the untrustworthy network traffic. 
Moreover, defenders can use intrusion detection systems \cite{axelsson2000intrusion} to detect a suspected malicious activity when an intrusion penetrates the system. 
These defensive methods are useful in deterring naive attacks up to certain degree. However, the unequal status between the attacker and the defender naturally gives the attacker an advantage in the game. An attacker succeeds by knowing and exploiting one zero-day vulnerability while a defender can be successful only when he can defend against all attacks. 
Moreover, attacks evolve to be increasingly sophisticated and can easily challenge the traditional defense mechanisms, i.e., intrusion prevention, detection, and response. 

Cyber deception is one way to evade the detection. 
As defined in \cite{sep-lying-definition}, the deception is either the prevention from a true belief or a formulation of a false belief. 
In the cybersecurity setting, the first type of deception corresponds to a stealthy attack where the virus can behave to be legitimate apparently and remain undetected.  For example, if a strategic attacker knows the pre-defined rules of the firewalls or the rule-based intrusion detection system, they can adapt their behaviors to avoid triggering the alarm. 
In the second type, for example, hackers can launch ``sacrificial attacks" to trick the defender into a false belief that all viruses have been detected and repelled \cite{TegraX1}.
The adversarial cyber deception introduces the information asymmetry and poses attackers in a favorable position.  A defender is analogous to a blind person who competes with a sighted attacker in a well-illuminated room. 

To tilt the information asymmetry, the defender can be reactive, i.e., continuously consummates the intrusion prevention and detection system capable of stealthy and deceptive attacks. This costly method is analogous to curing the blindness. 
Defensive deception, however, provides an alternative to the costly rectifications of the system by deliberately and proactively introducing uncertainties into the system, i.e., private information unknown to the attacker. This proactive method is analogous to turning off the light and providing every participant, especially the attacker with sufficient amount of uncertainties. 
For example, a system can include honeypots that contain no information or resource of value for the attackers. However, the defender can make the honeypot indistinguishable from the real systems by faking communication and network traffic. Since a legitimate user should not access the honeypot, the activities in the honeypot reveal the existence as well as characteristics of that attack.

The cyber attacks and defenses are the spear and shield, the existence of attackers motivates the development of defensive technologies, which in turn stimulates advanced attacks that are strategic, deceptive, and persistent. 
In this chapter, we model these competitive interactions using game theory ranging from complete to incomplete information, static to multi-stage transition, and symmetric to asymmetric information structures. 

\subsection{Literature}
Deception and its modeling are emerging areas of research.
The survey \cite{pawlick2017game} provides a taxonomy that defines six types of defensive deception: perturbation via external noises, moving target defense (MTD), obfuscation via revealing useless information, mixing via exchange systems, honey-$x$, and the attacker engagement that uses feedback to influence attackers dynamically. 
MTD \cite{jajodia2011moving} can limit the effectiveness of the attacker's reconnaissance by manipulating the attack surface of the network. The authors
in  \cite{zhu2013game} combine information- and control-theory to design an optimal MTD mechanism based on a feedback information structure while
\cite{maleki2016markov}, \cite{lei2017optimal} use the Markov chain to model the MTD process and discuss the optimal strategy to balance the defensive benefit and the network service quality. 

Game-theoretic models are natural frameworks to capture the adversarial and defensive interactions between players \cite{zhu2018multi,rass2017physical,zhuang2010modeling,miao2018hybrid,farhang2014dynamic,manshaei2013game,zhu2013deployment,zhang2017strategic,horak2017manipulating,huang2017large}. There are two perspectives to deal with the incomplete information under the game-theoretic setting, i.e., the robust game theory \cite{aghassi2006robust} that conservatively considers the worst case and the Bayesian game model \cite{harsanyi1967games} that introduces a random variable called the \textit{type} and the concept of Bayesian strategies and equilibrium. 
Signaling game, a two-stage game with one-sided incomplete information has been widely applied to different cybersecurity scenarios. For example, \cite{zhuang2010modeling} considers a multiple-period signaling game in the attacker-defender resource-allocation. The authors in \cite{pawlick2015deception} combine the signaling game with an external detector to provide probabilistic warnings when the sender acts deceptively. 
The recent work of \cite{huang2018gamesec} has proposed a multi-stage Bayesian game with two-sided incomplete information  that well characterizes the composite attacks that are advanced, persistent, deceptive and adaptive. A dynamic belief update and long-term statistical optimal defensive policies are proposed to mitigate the loss and deter the adversarial users.


\subsection{Notation}
In this chapter, the pronoun `he' refers to the user denoted by $P_2$, and `she' refers to the defender as $P_1$. Calligraphic fonts such as $\mathcal{A}$ represent a set. 
For $i\in \mathcal{I}$, notation `$-i$' means $\mathcal{I}\setminus \{i\}$. 
Take  $\mathcal{I}:=\{1,2\}$ as an example, if $i=1$, then $-i=2$. 
If $\mathcal{A}$ is a finite set, then we let $\bigtriangleup \mathcal{A}$ represent the set of probability distributions over $\mathcal{A}$, i.e., $\bigtriangleup \mathcal{A}:=\{p:\mathcal{A} \mapsto R_{+} | \sum_{a\in \mathcal{A}} p(a)=1\}$.

\section{Static Game with Complete Information  for Cybersecurity}
Game theory has been applied to cybersecurity problems \cite{pawlick2017proactive,chen2018security,zhang2017strategic,manshaei2013game,pawlick_mean-field_2017,xu_game-theoretic_2017,pawlick2018modeling} to  capture quantitatively the interaction between different ``players" including the system operator, legitimate users, and malicious hackers. 
As a baseline security game, the bi-matrix game focuses on two non-cooperative players, i.e., an attacker $P_2$ aiming at compromising the system and a defender $P_1$ who tries to prevent systems from adverse consequences, mitigate the loss under attacks, and recover quickly and thoroughly to the normal operation after the virus' removal.

Each player $P_i, i\in \{1,2\}$ can choose an action $a_i$ from a finite set $\mathcal{A}_i$ and $m_i:=|\mathcal{A}_i|$ is the number of actions $P_i$ can choose from. 
The value of the utility $J_i(a_1,a_2)\in \mathcal{R}^{m_1\times m_2}$ for each player $i$ depends collectively on both players' actions as shown in Table \ref{table: static game}. 
As stated in the introduction, targeted attacks can investigate the system thoroughly,  exploit vulnerabilities, and obtain the information on the security settings including the value of assets and possible defensive actions. 
Thus, the baseline game with complete information assumes that both players are aware of the other player's existence, action sets, and payoff matrices. However, each player will not know the other player's action before making his/her decision. 
Example \ref{ex: static game} considers a nonzero-sum complete-information security game where the attacker and the defender have conflicting objectives, i.e.,  $\exists a_1\in \mathcal{A}_1, a_2\in \mathcal{A}_2, J_1(a_1,a_2)+J_2(a_1,a_2) \neq 0$.
For scenarios where the defender does not know the utility of the attacker, she can assume $J_2(a_1,a_2) =-J_1(a_1,a_2), \forall a_1\in \mathcal{A}_1, a_2\in \mathcal{A}_2$ and use the zero-sum game to provide a useful worst-case analysis.

\begin{table}[h]
\centering
\caption{Utility bi-matrix $(J_1,J_2)$ of the static secure game, i.e.,  $J_1=[0,-r_1;0,r_3], J_2=[0,r_2;0, -r_4]$. $P_1$ is the row player and $P_2$ is the column player. Both players are rational and aim to maximize their own payoffs.
}
\label{table: static game}
\begin{tabular}{|l|l|l|l|}
\hline 
$P_1$ $\setminus$ $P_2$  &   NOP & Escalate\\ \hline
Permit               & $(0,0)$   & $(-r_1, r_2)$    \\ \hline
Restrict          & $(0,0)$  & $(r_3,-r_4)$   \\ \hline
\end{tabular}
\end{table}

\begin{example}
\label{ex: static game}
Consider the game in Table \ref{table: static game}. Attacker $P_2$ can either choose action $a_2=1$ to escalate his privilege in accessing the system, or choose No Operation Performed (NOP) $a_2=0$. Defender $P_1$ can either choose to restrict $a_1=1$ or allow $a_1=0$ a privilege escalation. The value in the brackets $(\cdot,\cdot)$ represents the utility for $P_1,P_2$ under the corresponding action pair, e.g., if the attacker escalates his privilege and the defender chooses to allow an escalation, then $P_2$ obtains  a reward of $r_2>0$ and $P_1$ receives a loss of $r_1>0$. 
In this example, no dominant (pure)-strategies exist for both players to maximize their utilities, i.e., each player's optimal action choice depends on the other player's choice. For example, $P_1$ prefers to allow an  escalation only when $P_2$ chooses the action NOP; otherwise $P_1$ prefers to restrict an escalation. 
The above observation motivates the introduction of the mixed-strategy in Definition \ref{def: mixed-strategy} and  the concept of Nash equilibrium in Definition \ref{def: NE} where any unilateral deviation from the equilibrium does not benefit the deviating player. 
\qed 
\end{example}

\begin{definition}
\label{def: mixed-strategy}
A mixed-strategy $\sigma_i\in \bigtriangleup \mathcal{A}_i$ for $P_i$ is a probability distribution on his/her action set $\mathcal{A}_i$. 
\qed
\end{definition}
Denote $\sigma_i(a_i)$ as $P_i$'s probability of taking action $a_i$, then $\sum_{a_i\in \mathcal{A}_i}\sigma_i(a_i)=1,\forall i\in \{1,2\}$ and $\sigma_i(a_i)\geq 0, \forall i\in \{1,2\}, a_i\in A_i$. Once player $P_i$ has determined strategy $\sigma_i$, the action $a_i$ will be a realization of the strategy. Hence, each player $P_i$ under the mixed-strategy has the objective to maximize the expected utility $\sum_{a_1\in \mathcal{A}_1}\sum_{a_2\in \mathcal{A}_2}\sigma_1(a_1)\sigma_2(a_2)J_1(a_1,a_2)$. 
Note that the concept of the mixed strategy includes the pure strategy as a degenerate case. 

\begin{definition}
\label{def: NE}
A pair of mixed-strategy $(\sigma^*_1,\sigma^*_2)$ is said to constitute a (mixed-strategy) Nash equilibrium (NE) if for all $\sigma_1\in \bigtriangleup \mathcal{A}_1, \sigma_2\in \bigtriangleup \mathcal{A}_2$, 
\begin{align*}
\sum_{a_1\in \mathcal{A}_1}\sum_{a_2\in \mathcal{A}_2}\sigma^*_1(a_1)\sigma^*_2(a_2)J_1(a_1,a_2)\geq \sum_{a_1\in \mathcal{A}_1}\sum_{a_2\in \mathcal{A}_2}\sigma_1(a_1)\sigma^*_2(a_2)J_1(a_1,a_2),\\
\sum_{a_1\in \mathcal{A}_1}\sum_{a_2\in \mathcal{A}_2}\sigma^*_1(a_1)\sigma^*_2(a_2)J_1(a_1,a_2)\geq \sum_{a_1\in \mathcal{A}_1}\sum_{a_2\in \mathcal{A}_2}\sigma^*_1(a_1)\sigma_2(a_2)J_1(a_1,a_2).
\end{align*}
\qed
\end{definition}
In a finite static game with complete information, the mixed-strategy Nash equilibrium always exists. Thus, we can compute the equilibrium which may not be unique via the following system of equations.
\begin{align*}
\sigma_1^* \in  arg \max_{\sigma_1} \sum_{a_1\in \mathcal{A}_1}\sum_{a_2\in \mathcal{A}_2}\sigma_1(a_1)\sigma^*_2(a_2)J_1(a_1,a_2),\\
\sigma_2^* \in  arg \max_{\sigma_2} \sum_{a_1\in \mathcal{A}_1}\sum_{a_2\in \mathcal{A}_2}\sigma^*_1(a_1)\sigma_2(a_2)J_1(a_1,a_2).
\end{align*}
The static game model and equilibrium analysis are useful in the cybersecurity setting because of the following reasons. 
First, the strategic model quantitatively captures the competitive interaction between the hacker and the system defender. 
Second, the NE provides a prediction of the security outcomes of the scenario which the game model captures. 
Third, the probabilistic defenses suppress the probability of adversarial actions and thus mitigate the expected economic loss. 
Finally, the analysis of the equilibrium motivates an optimal security mechanism design which can shift the equilibrium toward ones that are favored by the defender via an elaborate design of the game structure.

\section{Static Games with Incomplete Information for Cyber Deception}
\label{Sec: staticB}
The primary restrictive assumption for the baseline security game is that all game settings including the action sets and the payoff matrices are of complete information to the players. However, the deceptive and stealthy nature of advanced attackers makes it challenging for the defender to identify the nature of the malware accurately at all time. Even the up-to-date intrusion detection system has the false alarms and misses that can be fully characterized by a receiver operating characteristic (ROC) curve plotted with the true positive rate (TPR) against the false positive rate (FPR).
To capture the uncertainty caused by the cyber deception, we introduce a random variable called the \textit{type} to model the possible scenario variations as shown in Example \ref{ex: static Bayesian}.
\begin{table}[h]
\centering
\caption{Utility bi-matrix when user $P_2$ is either adversarial $\theta_2=\theta^b$ or legitimate $\theta_2=\theta^g$.
}
\label{table: static Bayesian}
\begin{tabular}{|l|l|l|l|}
\hline 
$\theta_2=\theta^b$  &   NOP & Escalate\\ \hline
Permit               & $(0,0)$   & $(-r_2, r_2)$    \\ \hline
Restrict          & $(0,0)$  & $(r_0,-r_0)$   \\ \hline
\end{tabular}
\quad
\begin{tabular}{|l|l|l|l|}
\hline 
$\theta_2=\theta^g$   &   NOP & Escalate\\ \hline
Permit               & $(0,0)$   & $(r_1, r_1)$    \\ \hline
Restrict          & $(0,0)$  & $(-r_1,-r_1)$   \\ \hline
\end{tabular}
\end{table}
\begin{example}
\label{ex: static Bayesian}
Consider the following static Bayesian game where
we use two discrete values of the type $\theta_2\in \Theta_2:=\{\theta^b,\theta^g\}$ to distinguish the user $P_2$ as either an attacker $\theta_2=\theta^b$ or a legitimate user $\theta_2=\theta^g$. 
The attacker can camouflage to be a legitimate user and possess the same action set $\mathcal{A}_2$, e.g., both attacker and legitimate can request to escalate the privilege $a_2=1$. 
However, since they are of different types, the introduced utilities $\bar{J}_i (a_1,a_2, \theta_2), i\in \{1,2\}$ are different under the same action pair$(a_1,a_2)$ as shown in Table \ref{table: static Bayesian}. 
For example, the privilege escalation has a positive effect on the system when the user $P_2$ is legitimate, yet will harm the system when $P_2$ is an attacker. 
Since the defender does not know the type of the user due to the cyber deception, we extend the Nash equilibrium analysis of the complete-information game to Bayesian Nash equilibrium in Definition \ref{def: BNE} to deal with the type uncertainty. 
Since  $P_2$ knows his type value to be either $\theta^g$ or $\theta^b$, his mixed-strategy $\bar{\sigma}_2: \Theta_2 \mapsto \bigtriangleup \mathcal{A}_2$ should be a function of his type value. Thus, with a slight abuse of notation, $\bar{\sigma}_2(a_2,\theta_2)\geq 0, \forall a_2\in \mathcal{A}_2, \forall \theta_2\in \Theta_2$ is the probability of taking action $a_2$ under the type value $\theta_2$. Clearly, the mixed-strategy is a probability measure, i.e.,  $\sum_{a_2\in \mathcal{A}_2} \bar{\sigma}_2(a_2,\theta_2)=1, \forall \theta_2\in \Theta_2$.
Suppose that $P_1$ manages to know the probability distribution of the type $b^0_1 \in \bigtriangleup \Theta_2$, e.g., defender $P_1$ believes with probability $b^0_1(\theta^g)$ that user $P_2$ is of a legitimate type and $b^0_1(\theta^b)$ that  $P_2$ is of an adversarial type. Similarly, we have $\sum_{\theta_2\in \Theta_2} b^0_1(\theta_2)=1$ and $b^0_1(\theta_2)\geq 0, \forall \theta_2\in \Theta_2$.
\qed
\end{example} 

\begin{definition}
\label{def: BNE}
A pair of mixed-strategy $(\sigma^*_1,\bar{\sigma}_2^*(\cdot))$ is said to constitute a (one-sided) mixed-strategy Bayesian Nash equilibrium (BNE) if 
\begin{align*}
& \sum_{\theta_2\in \Theta_2} b^0_1(\theta_2) \sum_{a_1\in \mathcal{A}_1}\sum_{a_2\in \mathcal{A}_2}\sigma^*_1(a_1)\bar{\sigma}^*_2(a_2, \theta_2)\bar{J}_1(a_1,a_2, \theta_2)
 \geq  \\
&  \sum_{\theta_2\in \Theta_2} b^0_1(\theta_2) \sum_{a_1\in \mathcal{A}_1}\sum_{a_2\in \mathcal{A}_2}\sigma_1(a_1)\bar{\sigma}^*_2(a_2, \theta_2)\bar{J}_1(a_1,a_2, \theta_2), \forall \sigma_1(\cdot).
\end{align*}
and 
\begin{align*}
&  \sum_{a_1\in \mathcal{A}_1}\sum_{a_2\in \mathcal{A}_2}\sigma^*_1(a_1)\bar{\sigma}^*_2(a_2, \theta_2)\bar{J}_2(a_1,a_2, \theta_2)
 \geq  \\
&  \sum_{a_1\in \mathcal{A}_1}\sum_{a_2\in \mathcal{A}_2}\sigma^*_1(a_1)\bar{\sigma}_2(a_2, \theta_2)\bar{J}_2(a_1,a_2, \theta_2), \forall \theta_2\in \Theta_2, \forall \bar{\sigma}_2(\cdot, \theta_2).
\end{align*}
\qed
\end{definition}
Note that the binary type space $\Theta_2$ can easily extend to finitely many elements to model different kinds of legitimate users and hackers who bear diverse type-related payoff functions. 
Since the type distinguishes different users and characterizes their essential attributes, the type space can also be a continuum and interpreted as a normalized measure of damages or the threat level to the system \cite{huang2018PER}. 
Moreover, the defender $P_1$ can also have a type $\theta_1\in \Theta_1$, which forms a static version of the two-sided dynamic Bayesian game as shown in Section \ref{Sec: TIFS}.  
Theorem \ref{thm: BNE} guarantees the existence of BNE regardless of extensions mentioned above. 

\begin{theorem}
A mixed-strategy BNE exists for a static Bayesian game with a finite type space. For games with a continuous type space and a continuous strategy space, if strategy sets and type sets are compact, payoff functions are continuous and concave in players' own strategies, then a pure-strategy BNE exists. 
\label{thm: BNE}
\end{theorem}

\section{Dynamic Bayesian Game  for Deception and Counter-Deception}
Followed from the above static Bayesian game with one-sided incomplete information, we investigate two types of dynamic games for cyber deception and counter-deception.
The signaling game is two-stage and only the receiver has the incomplete information of the sender's type. 
The two-sided dynamic Bayesian game
with a multi-stage state transition in Section \ref{Sec: TIFS} can be viewed as an extension of the signaling game.
The solution concept in this section extends the BNE to the perfect Bayesian Nash equilibrium (PBNE). 

\subsection{Signaling Game for Cyber Deception}
We illustrate the procedure of the signaling game as follows:
\begin{itemize}
\item An external player called the Nature draws a type $\theta_2$ from a set $\Theta_2:=\{\theta^1,\theta^2,\cdots, \theta^I \}$ according to a given probability distribution $b^0_1\in \bigtriangleup \Theta_2$ where $b^0_1(\theta^i)\geq 0, \forall i\in \{1,2,\cdots,I\}$ and $\sum_{i=1}^I b^0_1(\theta^i)=1$. 
\item The user $P_2$ (called the sender) observes the type value $\theta_2$ and then chooses an action $a_2$ (called a message) from a finite set of message space $\mathcal{A}_2$.
\item The defender $P_1$ (called the receiver) observes the action $a_2$ and then chooses her action $a_1\in \mathcal{A}_1$. 
\item Payoffs $(\bar{J}_1(a_1,a_2,\theta_2),\bar{J}_2(a_1,a_2,\theta_2))$ are given to the sender and receiver, respectively.
\end{itemize}
\paragraph{Belief Formulation.}
Since the receiver $P_1$ has incomplete information about the sender's type, she will form a belief $b_1^1: \mathcal{A}_2 \mapsto \bigtriangleup \Theta_2$ on the type $\theta_2$ based on the observation of the sender's message $a_2$. As a measure of the conditional probability, the belief $b_1^1$ satisfies
$b^1_1(\theta^i | a_2 )\geq 0, \forall i\in \{1,2,\cdots,I\},\forall a_2\in \mathcal{A}_2$ and $\sum_{\theta_2\in \Theta_2}b_1^1(\theta_2|a_2):=\sum_{i=1}^I b^1_1(\theta^i | a_2)=1, \forall a_2\in \mathcal{A}_2$. 

\paragraph{Receiver's Problem.} For every received message $a_2$, receiver $P_1$ aims to optimize her expected payoffs under her belief $b_1^1(\cdot|a_2)$, i.e., 
\begin{align}
\label{eq:pureReceiver}
\max_{a_1\in \mathcal{A}_1} \sum_{\theta_2\in \Theta_2}  b_1^1(\theta_2|a_2)\bar{J}_1(a_1,a_2,\theta_2). 
\end{align}
As a result, the receiver's (pure)-strategy is given by the mapping $\hat{a}_1: \mathcal{A}_2 \mapsto \mathcal{A}_1$. Thus, the receive $P_1$'s  action is the outcome of the mapping, i.e., $a_1=\hat{a}_1(a_2)$. 
\paragraph{Sender's Problem.}
For every type $\theta_2\in \Theta_2$ that the Nature picks for $P_2$, sender $P_2$ should pick a message $a_2\in \mathcal{A}_2$ that maximizes the following utility with the anticipation of receiver's action $a_1=\hat{a}_1(a_2)$, i.e., 
\begin{align}
\label{eq:pureSender}
\max_{a_2\in \mathcal{A}_2} \bar{J}_2(\hat{a}_1(a_2),a_2, \theta_2).
\end{align}
Hence, the sender's (pure)-strategy is given by the mapping $\bar{a}_2: \Theta_2 \mapsto \mathcal{A}_2$ and $P_2$'s action under the type value $\theta_2$ is $a_2=\bar{a}_2(\theta_2)$. 
The sender's strategy $\bar{a}_2$ is called a pooling strategy if he chooses the same message $a_2$ independent of the type given by the Nature, and is called a separating strategy if the mapping $\bar{a}_2$ is injective. For all other feasible mappings, $\bar{a}_2$ is called a semi-separating strategy. 

\paragraph{Mixed-strategy Receiver and Sender's Problem.} 
We can extend the pure-strategy to the mixed-strategy $\hat{\sigma}_1: \mathcal{A}_2 \mapsto \bigtriangleup \mathcal{A}_1$ for receiver $P_1$ and the same $\bar{\sigma}_2: \Theta_2 \mapsto \bigtriangleup \mathcal{A}_2$ defined in Section \ref{Sec: staticB} for sender $P_2$.  
After observing sender's message $a_2$ as a realization of the mix-strategy $\bar{\sigma}_2$, receiver $P_1$ assigns probability $\hat{\sigma}_1(a_1, a_2)$ to her action $a_1$ with the feasibility constraint $\sum_{a_1\in \mathcal{A}_1} \hat{\sigma}_1(a_1, a_2)=1, \forall a_2\in \mathcal{A}_2$ and  $\hat{\sigma}_1(a_1, a_2)\geq 0, \forall a_1\in \mathcal{A}_1, a_2\in \mathcal{A}_2$. 
The expected objective functions for both players under the mixed-strategy are defined as follows.
\begin{equation}
\label{eq: mixedSignaling}
\begin{split}
&\max_{\hat{\sigma}_1(\cdot)} 
\sum_{\theta_2\in \Theta_2}  b_1^1(\theta_2|a_2)
\sum_{a_1\in \mathcal{A}_1} \hat{\sigma}_1(a_1, a_2)
\bar{J}_1(a_1,a_2,\theta_2), \forall a_2\in \mathcal{A}_2. 
\\
&\max_{\bar{\sigma}_2(\cdot)} 
\sum_{a_1\in \mathcal{A}_1} \hat{\sigma}_1(a_1, a_2)\sum_{a_2\in \mathcal{A}_2}\bar{\sigma}_2(a_2, \theta_2)
\bar{J}_2(a_1, a_2, \theta_2), \forall \theta_2\in \Theta_2.
\end{split}
\end{equation}

\paragraph{Belief Consistency. }
Since the message $a_2$ is a function of the type $\theta_2$, the observation of the message should reveal some information of the type. Thus, the receiver updates the initial belief $b_1^0(\cdot)$ to form the posterior belief $b_1^1(\cdot|a_2)$ via the Bayesian rule. 
\begin{equation}
\label{eq: beliefSign}
\begin{split}
&b_1^1(\theta_2|a_2)=\frac{b_1^0(\theta_2)\bar{\sigma}_2(a_2|\theta_2)}{\sum_{\theta_2\in \Theta_2} b_1^0(\theta_2)\bar{\sigma}_2(a_2|\theta_2)}, \text{ if } \sum_{\theta_2\in \Theta_2} b_1^0(\theta_2)\bar{\sigma}_2(a_2|\theta_2)>0, \\
&b_1^1(\theta_2|a_2)=\text{any proability distritbutions}, \text{ if } \sum_{\theta_2\in \Theta_2} b_1^0(\theta_2)\bar{\sigma}_2(a_2|\theta_2)=0. 
\end{split}
\end{equation}
Serving as a particular case, the receiver and the sender's problem under the pure strategy should also satisfy the Bayesian update of the belief. 
Note that although $P_1$ can observe the message $a_2$ which is a realization of $\bar{\sigma}_2$, she cannot directly update her belief via \eqref{eq: beliefSign} if the signaling game is only played once. However, \eqref{eq: beliefSign} contributes to the PBNE of the signaling game in Definition \ref{def: PBNE}, serving as the belief consistency constraint.

\begin{definition}
\label{def: PBNE}
A pure-strategy perfect Bayesian Nash equilibrium of the signaling game is a pair of strategies $(\hat{a}^*_1,\bar{a}^*_2)$ and belief $b_1^{1,*}$ that satisfy \eqref{eq:pureReceiver} \eqref{eq:pureSender} and  \eqref{eq: beliefSign}. A mixed-strategy perfect Bayesian Nash equilibrium of the signaling game is a pair of strategies $(\hat{\sigma}^*_1,\bar{\sigma}^*_2)$ and belief $b_1^{1,*}$ that satisfy \eqref{eq: mixedSignaling} and \eqref{eq: beliefSign}.
\qed
\end{definition}
The reader may already realize that we can use signaling game to model the same cyber deception scenario in Example \ref{ex: static Bayesian} only with the difference of the asymmetric \textbf{information structure}, i.e.,  the defender $P_1$ has a chance to observe the behavior of the user $P_2$ before making her decision. The information asymmetry results in the following changes. 
First, $P_1$'s mixed-strategy $\hat{\sigma}_1(a_2)$ is a function of her observation, i.e., $P_2$'s action $a_2$. 
Second, instead of directly taking an expectation over the initial belief $b_1^0$, defender $P_1$ obtains a posterior belief $b_1^1$ that is consistent with the new observation $a_2$. 
Third, the type of belief  can affect the PBNE 
even under the cheap-talk setting when utilities of both players are independent of the message. 
Finally, if there is only one type with a known $b_1^0$, which means that the type value becomes common knowledge, the signaling game becomes a Stackelberg game with leader $P_2$ and follower $P_1$. 

\subsection{Multi-stage with Two-sided Incomplete Information}
\label{Sec: TIFS}
The deceptive techniques adopted by the attacker make it challenging for the defender to correctly identify the type of the user even observing the manifested behavior as shown in Example \ref{ex: static Bayesian}. 
To tilt the information asymmetry, we can either continue to develop the intrusion detection system to increase the TPR with decreased FPR or refer to defensive deception techniques to create a sufficient amount of uncertainties for the attackers. 
Use defensive and active deception as a counter-deception technique will disorient and slow down the adversarial infiltration because attackers have to judge the target's type, i.e., whether it is a real valuable production system or a well-pretended honeypot. 
Therefore, we introduce a two-sided incomplete information Bayesian game model with a multistage state transition for advanced attacks such as Advanced Persistent Threats (APTs) which infiltrate stage by stage. 
 
\subsubsection{Two-sided Private Types}
\label{sec: two-sided private tyep}
This section discusses the scenarios where not only the user $P_2$ has a type, the defender $P_1$ also has a private type $\theta_1\in \Theta_1$ to distinguish a system's different levels of sophistication and security awareness. 
For example, the defender's type space can be binary $\Theta_1:=\{\theta^H,\theta^L\}$ where $\theta^H$ represents a  defender who is well-trained with a high-security awareness and also supported by advanced virus detection and analysis systems. 
Thus, she may refer to the log file with a higher frequency and more likely to obtain valuable information through the behavior analysis. 
Thus, once the attacker requests for privilege escalation and $P_1$ restricts and inspects the log file, a higher reward as well as  a higher penalty are introduced under a high-type defender $\theta^H$ than a low-type defender $\theta^L$, i.e., $r_0=r_3 \cdot \mathbf{1}_{\theta^L}+r_4 \cdot \mathbf{1}_{\theta^H}$ where $r_4>r_3>0$ as shown in Table \ref{table: muti-stageBay}. 
\begin{table}[h]
\centering
\caption{Utility bi-matrix when user $P_2$ is either adversarial $\theta_2=\theta^b$ or legitimate $\theta_2=\theta^g$ and defender $P_1$ is either of high type $\theta_1=\theta^H$ or of low type $\theta_1=\theta^L$. 
}
\label{table: muti-stageBay}
\begin{tabular}{|l|l|l|l|}
\hline 
$\theta_2=\theta^b$  &   NOP & Escalate\\ \hline
Permit               & $(0,0)$   & $(-r_2, r_2)$    \\ \hline
Restrict          & $(0,0)$  & $(r_0,-r_0)$   \\ \hline
\end{tabular}
\quad
\begin{tabular}{|l|l|l|l|}
\hline 
$\theta_2=\theta^g$   &   NOP & Escalate\\ \hline
Permit               & $(0,0)$   & $(r_1, r_1)$    \\ \hline
Restrict          & $(0,0)$  & $(-r_1,-r_1)$   \\ \hline
\end{tabular}
\end{table}

Two aspects motivate us to introduce a random variable as the defender's type, i.e., the user $P_2$ only knows the prior probability distribution over the type space $\Theta_1$ yet not the value/realization of $P_1$'s type.
On the one hand, 
the modern cyberinfrastructure networks have become increasingly interdependent and complicated, so it is hard to evaluate the system payoff accurately even given both players' actions. 
On the other hand, 
the adoption of defensive deception techniques brings uncertainties and difficulties for the user, especially attackers to evaluate the system setting.
Therefore, we model the uncertainties by letting the utility function be a function of the type, which is a random variable. 

\subsubsection{A Scenario of Advanced Persistent Threats }
\begin{figure}[h]
  \centering
  \includegraphics[width=1\columnwidth]{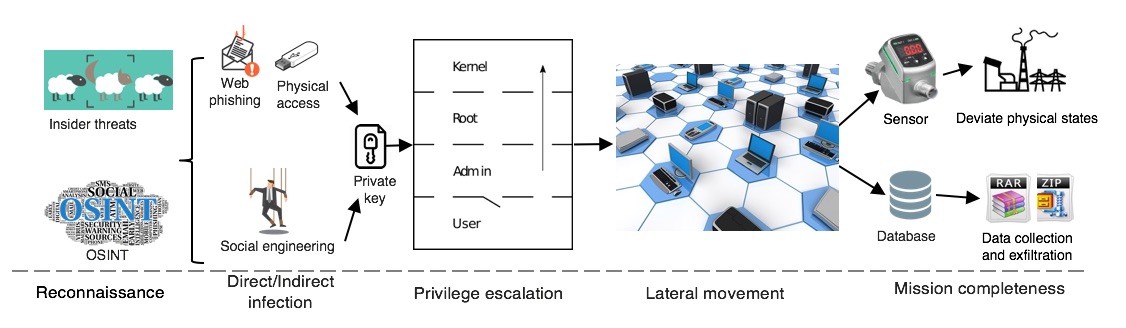}
  \caption{\label{attack graph} The multi-stage life cycle of APTs forms a tree structure. 
  During the reconnaissance phase, the threat actor probes the system and obtains intelligence from open-source information or insiders. The infection can be either directly through the web phishing and the physical access or indirectly through social engineering to manipulate the employees and then obtain a private key. 
Then, APTs gain the foothold, escalate privilege, propagate laterally in the cyber network and finally either cause physical damages or collect confidential data.  }
\end{figure}

A class of stealthy and well-planned sequence of hacking processes called Advanced Persistent Threats (APTs) motivates the multi-stage transition as well as two strategic players with two-sided incomplete information \cite{zhu2018multi,huang2018gamesec}. 
Unlike the non-targeted attacks who spray a large number of phishing emails and pray for some phools to click on the malicious links and get compromised, nation-sponsored APTs have sufficient amount of resources to initiate a reconnaissance phase to understand their targeted system thoroughly and tailor their attack strategies with the target.
Multistage movement is an inherent feature of APTs as shown in Fig. \ref{attack graph}.
The APTs' life cycle includes a sequence of stages such as the initial entry, foothold establishment, privilege escalation, lateral movement, and the final targeted attacks on either confidential data or the physical infrastructures such as nuclear power stations and automated factories. APTs use each stage as a stepping stone for the next one. 
Unlike the static ``smash and grab"  attacks who launch direct attacks to obtain one-shot reward and then get identified and removed, APTs possess a long-term persistence and stage-by-stage infiltration to evade detection. 
For example,  APTs can stealthily scan the port slowly to avoid hitting the warning threshold of the IDS. 
APTs hide and behave like legitimate users during the escalation and prorogation phases to deceive the defender until reaching the final stage, launch a `critical hit' on their specific targets, and cause an enormous loss.

The classical intrusion prevention (IP) techniques such as the cryptography and the physical isolation can be ineffective for APTs because APTs can steal full cryptographic keys by techniques such as social engineering. 
Stuxnet, as one of the most well-known APTs, has proven to be able to successfully bridge the air gap between local area networks with the insertion of infected USB drives.
Similarly, the intrusion detection (ID) approach including  \cite{coppolino2010intrusion} can be ineffective if APTs acquire the setting of the detection system with the help of insiders. Moreover, APTs operated by human-expert can analyze system responses and learn the detection rule during their inactivity, thus deceive the system defender and evade detection.
Additionally, 
APTs can encrypt the data as well as their communication content with their human experts. A well-encrypted outbound network flow will limit the effectiveness of the data loss prevention (DLS) system which detects potential data ex-filtration transmissions and prevents them by monitoring, detecting, and blocking sensitive data.

Hence, besides traditional defensive methods, i.e., IP, ID, DLS, 
it is essential to design strategic security mechanisms to capture the competitive  interaction, the multi-stage multi-phase transition, as well as the adversarial and defensive deception between the APTs and advanced defenders. 

As shown in Table \ref{table: muti-stageBay}, the advanced defender with a private type is deceptive and increases the attacker's uncertainty. The defender is also adaptive because she forms and updates the belief of the user's type according to the observation of the user's actions as shown in \eqref{eq: BayesianUpdate}.

 \subsubsection{Multi-stage State Transition}
 \label{sec: multi-stageBay}
As shown in Fig. \ref{attack graph}, the APT attacker moves stage by stage from the initial infection to the final target without jumps of multiple stages in one step. There are also no incentives for the attacker to go back to stages that he has already compromised because his ultimate goal is to compromise the specific target at the final stage. 
Therefore,  we model the APT transition as a multistage game with a finite horizon $K$. 
Each player $i\in \{1,2\}$ at each stage $k\in \{0,1,\cdots,K\}$ can choose an action $a_i^k$ from a stage-dependent finite set $\mathcal{A}_i^k$ because the feasible actions are different for each player at different stages. 
The history ${h}^k:=\{a_1^0,\cdots,a_1^{k-1},a_2^0,\cdots,a_2^{k-1}\} \in \mathcal{H}^k$ contains the actions of both players up to stage $k-1$ and can be obtained by reviewing system activities from the log file. 
Note that user's actions $a_2^k\in \mathcal{A}_2^k$ are the behaviors that are directly observable such as the privilege escalation request and the sensor access in the case study of Section \ref{sec: Casestudy}. 
Sine both legitimate and adversarial users can take these activities, a defender cannot identify the user's type directly from observing these actions.
On the other hand, the defender's action $a_1^k$ will be mitigation or proactive actions such as restricting the escalation request or monitoring the sensor access. These proactive actions also do not directly disclose the system type.  

State $x^k\in \mathcal{X}^k$ representing the status of the system at stage $t$ is the sufficient statistic of the history $\mathbf{h}^k$ because a Markov state transition $x^{k+1}=f^k(x^k,a_1^k,a_2^k)$ contains all the information of the history update $\mathbf{h}^k=\mathbf{h}^{k-1} \cup \{a_1^k,a_2^k\}$. 
Unlike the history, the cardinality of the state does not necessarily grows with the number of stages.
The function $f^k$ is deterministic because history is fully observable without uncertainties. The function $f^k$  is also stage-dependent and represents different meanings. 
For example, in Section \ref{sec: Casestudy}, the state at the second last stage represents the current privilege level, while at the final stage,  the state indicates which sensors have been compromised.

\subsubsection{Behavior Mixed-strategy and Believe Update}
According to the information available at stage $k$, i.e., history $h^k$ and his/her type $\theta_i$, player $i$ takes a behavioral mixed-strategy ${\sigma}^k_i: \mathcal{H}^k \times \Theta_i \mapsto \bigtriangleup \mathcal{A}_i^k$ with the available information as the input of the function.
Note that $\sigma^k_i(a_i^k|{h}^k,\theta_i)\in \Sigma_i^k:=\{\sigma^k_i(a_i^k|{h}^k,\theta_i)\geq 0: \sum_{a_i^k \in \mathcal{A}_i^k}\sigma^k_i(a_i^k|{h}^k,\theta_i)=1\}$ is the probability of taking action $a_i^k$ given ${h}^k,\theta_i$ for all stage $k\in \{0,1,\cdots,K\}.$

To correspond to the challenge of incomplete information of the other player's type, each player $i$ forms a belief $b_i^k: \mathcal{H}^k \times \Theta_i \mapsto \bigtriangleup \Theta_{-i}$ that maps the available information $h^k,\theta_i$ to the distribution over the type space of the other player. Likewise, $b^k_i(\theta_{-i}|{h}^k, \theta_i)$ at stage $k$ is the conditional probability mass function (PMF) of the other player's type $\theta_{-i}$ and $\sum_{\theta_{-i}\in \Theta_{-i}} b^k_i(\theta_{-i}|{h}^k, \theta_i)d\theta_{-i}=1,  \forall k\in \{0,1,\cdots,K\}, \forall {h}^k \in \mathcal{H}^k ,\theta_i\in \Theta_i, i\in\{1,2\}$. 

Assume that each player $i$ knows the prior distribution of the other player's type, i.e., $b_i^0$ according to the historical data and the statistical analysis. If no prior information is available, a uniform distribution is an unbiased estimate. 
Since the multi-stage model provides a sequential observation of the other player's action $a_{-i}^k$ which is a realization of the mixed-strategy $\sigma_{-i}^k$, player $i$'s belief of the other's type can be updated via the Bayesian rule, i.e., 
\begin{equation}
\label{eq: BayesianUpdate}
\begin{split}
 {b}^{k+1}_i(\theta_{-i}|[{h}^k,{a}_i^k,{a}^k_{-i}], \theta_i)= \frac{{b}^k_i(\theta_{-i}|{h}^k, \theta_i)\sigma^k_{-i}({a}_{-i}^k|{h}^k,\theta_{-i})}{\sum_{{\theta}_{-i} \in {\Theta}_i } {b}^k_i({\theta}_{{-i}}|{h}^k, \theta_i)\sigma^k_{-i}({a}_{-i}^k|{h}^k,{\theta}_{-i})}.
\end{split}
\end{equation}

Note that the one-shot observation of the other player's action does not directly disclose the type because of the deception. However, since the utility function in Section \ref{sec: costfunction} is type dependent, the action made by the type-dependent policy will serve as a message that contributes to a better estimate of the other's type. The accuracy of the belief will be continuously improved when more actions are observed. 

\subsubsection{Utility Function and PBNE}
\label{sec: costfunction}
At each stage $k$,  $J_i^k$ is the utility that depends on the type and the action of both players, the current state ${x}^k$, and some external random noise $w_i^k$ with a known distribution. 
We introduce the external noise to model other unknown factors that could affect the value of the stage utility. The existence of the external noise makes it impossible for each player $i$ to directly acquire the value of the other's type $\theta_{-i}$ based on the combined observation of input parameters $x^k,a_1^k,a_2^k,\theta_i$ plus the output value of the utility function $J_i^k$. 
In the case study, we consider any additive noise with a $0$ mean, i.e., 
$
J_i^k(x^k,a_1^k,a_2^k, \theta_{i},\theta_{-i}, w_i^k)=\tilde{J}_i^k(x^k,a_1^k,a_2^k, \theta_{i},\theta_{-i})+w_i^k, 
$
which leads to an equivalent payoff over the expectation of the external noise $E_{w_i^k}J_i^k=\tilde{J}_i^k, \forall x^k,a_1^k,a_2^k, \theta_{i},\theta_{-i}$.

One significant improvement from the static game to the dynamic game is that each player $i$ has a long-term objective to maximize the total expected payoff $U_i^{k': K}$. 
For example, attackers of APTs may sacrifice the immediate attacking reward to remain stealthy and receive more considerable benefits in the following stages, e.g., successfully reach the final target and complete their mission. 
Define $\sigma_i^{k':K}:=[\sigma^k_i(a_i^k|{h}^{k},\theta_i)]_{k=k',\cdots,K}$ and the cumulative expected utility $U_i^{k':K}$ sums the expected stage utilities from stage $k'$ to $K$ as follows. 

\begin{equation}
\label{eq: cumultive utility}
\begin{split}
&U^{k':K}_i(\sigma_i^{k':K},\sigma_{-i}^{k':K}, {h}^{K+1},\theta_i) \\
&:=\sum_{k=k'}^{K} E_{\theta_{-i}\sim b_i^k, \sigma_i^k,\sigma_{-i}^k, w_i^k} [J_i^k({x}^k,\sigma_i^k,\sigma_{-i}^k,\theta_i,\theta_{-i},w_i^k)]\\
&=
\sum_{k=k'}^K \sum_{\theta_{-i}\in \Theta_{-i}} {b}^{k}_i(\theta_{-i}|{h}^k, \theta_i)     \sum_{a_{i}^k \in \mathcal{A}_{i}^k} \sigma_{i}^k(a_{i}^k |{h}^k, \theta_{i}) \\
& \cdot\sum_{a_{-i}^k \in \mathcal{A}_{-i}^k}\sigma_{-i}^k(a_{-i}^k |{h}^k, \theta_{-i}) \tilde{J}_i^k({x}^k,{a}_i^k,{a}_{-i}^k,\theta_{i},\theta_{-i}). 
\end{split}
\end{equation}

Similar to the PBNE of the signaling game, the PBNE of multi-stage Bayesian game defined in Definition \ref{def: multistage-PBNE} requires a $K$-stage belief consistency. Since the equilibrium may not always exist, an $\varepsilon$-equilibrium is introduced.  

\begin{definition}
\label{def: multistage-PBNE}
In the two-person $K$-stage Bayesian game with two-sided incomplete information and a cumulative payoff function $U_i^{k':K}$ in \eqref{eq: cumultive utility},  a sequence of strategies $\sigma_i^{*,k':K}\in \prod_{k=k'}^{K} \Sigma_i^{k}$ is called the $\varepsilon$ perfect Bayesian Nash equilibrium for player $i$, if $b_i^{k}$ satisfies the consistency constraint \eqref{eq: BayesianUpdate} for all $k\in \{0,1,\cdots,K-1\}$
 and for a given $\varepsilon\geq 0$, 
 \begin{align*}
U_1^{k:K}(\sigma_1^{*,k:K},\sigma_{2}^{*,k:K}, {h}^{K+1},\theta_1)\geq \sup_{\sigma_1^{k:K} } U_1^{k:K}(\sigma_1^{k:K},\sigma_{2}^{*,k:K}, {h}^{K+1},\theta_1)-\varepsilon. \\
U_2^{k:K}(\sigma_1^{*,k:K},\sigma_{2}^{*,k:K}, {h}^{K+1},\theta_2)\geq \sup_{\sigma_2^{k:K} } U_2^{k:K}(\sigma_{1}^{*,k:K}, \sigma_2^{k:K}, {h}^{K+1},\theta_2)-\varepsilon. 
\end{align*}
If $\varepsilon=0$, we have a perfect Bayesian Nash equilibrium.
\end{definition}

\subsubsection{Dynamic Programming}
Given any feasible belief at every stage, we can use dynamic programming to find the PBNE  in a backward fashion because of the tree structure and the finite horizon. 
Define the value function as the utility-to-go function under the PBNE strategy pair, i.e., 
\begin{align*}
V_i^{k}({h}^{k},\theta_i)=U_i^{k: K}(\sigma_i^{*,k:K},\sigma_{-i}^{*, k:K}, {h}^{k+1},\theta_i). 
\end{align*}
Let $V_i^{K+1}({h}^{K+1},\theta_i):=0$ be the boundary condition of the value function, we have the following recursive system equations to solve the PBNE mixed-strategies $\sigma_1^{*,k},\sigma_2^{*,k}$ for all stage $k=\{0,1,\cdots,K \}$: 

\begin{equation}
\label{eq: DP}
    \begin{cases} 
& V_1^{k-1}({h}^{k-1},\theta_1)= \sup_{\sigma_1^{k-1}\in \Sigma_1^{k-1}}  
 E_{\theta_{2}\sim b_1^{k-1},\sigma_1^{k-1},\sigma_{2}^{*,k-1}}\\
 & [ V_1^{k}([{h}^{k-1},a_1^{k-1},a_{2}^{k-1}],\theta_1) 
+  \tilde{J}_1^{k-1}(x^{k-1}, a_1^{k-1},a_{2}^{k-1},\theta_1,\theta_{2})]; \\
& V_2^{k-1}({h}^{k-1},\theta_2)=  \sup_{\sigma_2^{k-1}\in \Sigma_2^{k-1}} E_{\theta_{1}\sim b_2^{k-1},\sigma_{1}^{*,k-1},\sigma_2^{k-1}}  \\
&[ V_2^{k}([{h}^{k-1},a_1^{k-1},a_{2}^{k-1}],\theta_2) 
+  \tilde{J}_2^{k-1}(x^{k-1}, a_1^{k-1},a_{2}^{k-1},\theta_1,\theta_{2})]. \\
 \end{cases}
\end{equation}
Under the assumption of a Markov mixed-strategy $ \tilde{\sigma}_i^t(a_i^k|x^k,\theta_i)  \equiv {\sigma}_i^k(a_i^k|h^k,\theta_i)$, $\tilde{V}_i^k(x^k,\theta_i)$ becomes the sufficient statistics of $V_i^k(h^k,\theta_i)$. By replacing $\sigma_i^k(a_i^k|h^k,\theta_i)$ to $\tilde{\sigma}_i^k(a_i^k|x^k,\theta_i)$ and $V_i^k(h^k,\theta_i)$ to $\tilde{V}_i^k(x^k,\theta_i)$ in \eqref{eq: DP}, we can obtain a new  dynamic programming equation: 
\begin{equation}
\label{eq: DPsufficent}
\begin{split}
& \tilde{V}_i^{k-1}({x}^{k-1},\theta_i)= \sup_{\tilde{\sigma}_i^{k-1}}  
 E_{\theta_{-i}\sim b_i^{k-1},\tilde{\sigma}_i^{k-1},\tilde{\sigma}_{-i}^{*,k-1}}\\
 & [ \tilde{V}_i^{k}(f^k({x}^{k-1},a_1^{k-1},a_{2}^{k-1}),\theta_i) 
+  \tilde{J}_i^{k-1}(x^{k-1}, a_1^{k-1},a_{2}^{k-1},\theta_1,\theta_{2})].
\end{split}
\end{equation}

\subsubsection{PBNE Computation by Bilinear Programming}
To compute the PBNE, we need to solve a coupled system of the forward belief update in \eqref{eq: BayesianUpdate} that depends on the PBNE strategies plus a backward PBNE computation in \eqref{eq: DPsufficent} that can also be influenced by the type belief. 
If there are no additional structures to explore, we have to use a forward and backward iteration
with the boundary condition of the initial belief $b_i^0(\theta_{-i})$ and final stage utility-to-go $\tilde{V}_i^{K+1}(x^{K+1}, \theta_i)=0$. 
In particular, we first assign any feasible value to the type belief $b_i^k, k \in \{1,2\cdots,K\}$, then solve \eqref{eq: DPsufficent} from stage $k=K$ to $k=0$ and use the resulted PBNE strategy pair to update \eqref{eq: BayesianUpdate}. 
 We iteratively compute \eqref{eq: DPsufficent} and  \eqref{eq: BayesianUpdate} until both the $K$-stage belief and the PBNE strategy do not change, which provides a consistent pair of the PBNE and the belief. 
If the iteration process does not converge, then the PBNE does not exist. 
Define $l_{m_i}$ as the column vector of ones with a dimension of $m_i$, we propose a bilinear program to solve the PBNE strategy for any given belief $b_i^k, k \in \{1,2\cdots,K\}$, which leads to Theorem \ref{Thm: bilinear two-sided}. The type space can be either discrete or continuous. We refer reader to Section $4.4$ in \cite{huang2018gamesec} for the proof of the theorem.

\begin{theorem}\label{Thm: bilinear two-sided}
A strategy pair $(\tilde{\sigma}_1^{*,k},\tilde{\sigma}_2^{*,k})$ with the feasible state $x^k\in \mathcal{X}^k$ and the consistent belief sequence $b_i^k$ at stage $k\in \{0,1,\cdots,K\}$ constitutes a mixed-strategy PBNE of the multistage Bayesian game in Definition \ref{def: multistage-PBNE}, if, and only if, there exists a sequence of \textit{scalar function} pair $(s^{*,k}(\theta_1),w^{*,k}(\theta_2))$ such that $\tilde{\sigma}_1^{*,k}(\cdot|x^k,\theta_1),\tilde{\sigma}_2^{*,k}(\cdot|x^k,\theta_2)$, $s^{*,k}(\theta_1),w^{*,k}(\theta_2)$ are the optimal solutions to the following bi-linear program for each  $k\in \{0,1,\cdots,K\}$:

\begin{equation}
\label{eq: bilinear program}
\begin{split}
\sup_{\tilde{\sigma}^k_1,\tilde{\sigma}_2^k,s,w} \ & 
\sum_{\theta_1\in \Theta_1} b_1^k (\theta_1)
\sum_{\theta_2\in \Theta_2} b_2^k (\theta_2) \sum_{a_1^k\in \mathcal{A}_1^k}\tilde{\sigma}^k_1(a_1^k|x^k, \theta_1)\sum_{a_2^k\in \mathcal{A}_2^k}\tilde{\sigma}^k_2(a_2^k|x^k, \theta_2) \\
& \sum_{i=1}^2 [\tilde{J}_i^k(x^k,a_1^k,a_2^k, \theta_{1},\theta_{2}) +\tilde{V}_i^{k+1}(f^k(x^{k},a_1^k,a_2^k),\theta_i)]
\\
& +\sum_{\theta_2\in \Theta_2} b_2^k (\theta_2) w(\theta_2)+ \sum_{\theta_1\in \Theta_1} b_1^k (\theta_1) s(\theta_1)\\
s.t.    \quad  (a).  &
\sum_{\theta_1\in \Theta_1} b_1^k (\theta_1)\sum_{a_1^k\in \mathcal{A}_1^k}\tilde{\sigma}^k_1(a_1^k|x^k,\theta_1) [\tilde{J}_2^k(x^k,a_1^k,a_2^k, \theta_{1},\theta_{2})\\
& +\tilde{V}_2^{k+1}(f^k(x^{k},a_1^k,a_2^k),\theta_2)]
\leq -w(\theta_2)l_{m_2} , \forall \theta_2\in \Theta_2
\\
(b). &\sum_{\theta_2\in \Theta_2} b_2^k (\theta_2)\sum_{a_2^k\in \mathcal{A}_2^k}\tilde{\sigma}^k_2(a_2^k|x^k,\theta_2) [\tilde{J}_1^k(x^k,a_1^k,a_2^k, \theta_{1},\theta_{2})\\
& +\tilde{V}_1^{k+1}(f^k(x^{k},a_1^k,a_2^k),\theta_1)]
\leq -s(\theta_1) l_{m_1}, \forall \theta_1\in \Theta_1.
\end{split}
\end{equation}
\qed
\end{theorem}
Note that the solution of \eqref{eq: bilinear program} at stage $k+1$ provides the value of $\tilde{V}_i^{k+1}$ and $\tilde{V}_i^{K+1}=0$ is a known value. Thus, we can solve \eqref{eq: bilinear program} from $k=K$ to $k=0$ for any given type belief.

\subsubsection{An Illustrative Case Study}
\label{sec: Casestudy}
We adopt the same binary type space in Section \ref{sec: two-sided private tyep} and consider the following three-stage ($K=2$) transition. The proactive defensive actions listed in the case study should be combined with the reactive methods such as the firewall to defend attacks other than APTs. 

\paragraph{Initial Stage}
We consider the web phishing scenario for the initial entry. 
The state space $\mathcal{X}^0:=\{0,1\}$ of the initial stage is binary. Let $x^0=0$ represents that the user sends the email from an external IP domain while $x^0=1$ represents an email from the internal network domain. The attacker can also start from state $x^0=1$ due to the insider threats and the social engineering techniques. 

To penalize the adversarial exploitation of the open-source intelligence (OSINT) data, the defender can create avatars (fake personal profiles) on the social network or the company website. 
The user $P_2$ at the initial stage can send emails to a regular employee $a_2^0=0$, a Chief Executive Officer (CEO) $a_2^0=1$, or the avatar $a_2^0=2$. The email can contain a legitimate shortening Uniform Resource Locator (URL). 
  If the user is legitimate, the URL will lead to the right resources, yet if the user is malicious, the URL will redirect to a malicious site and then take control of the client's computer. 
As for the defender, suppose that $P_1$ proactively equips the computer with an anti-virus system that can run the email in the sandbox and apply penetration test. However, the limited budget can only support either the employees' computer or the CEO's computer. Thus, the defender also has three possible actions, i.e., equips the CEO' computer $a_1^0=2$, the employee's computer $a_1^0=1$, or does not equip the anti-virus system $a_1^0=0$ to avoid a deployment fee $c_0^0$. 
The defender of high-security awareness $\theta^H$ will deploy an advanced anti-virus system that costs higher installation fee than the regular anti-virus system, i.e.,  $c_2^0>c_1^0$, yet also provides a higher penalty to the attacker, i.e.,  $r_4^0>r_3^0$. Define $c_0^0:=c_1^0\mathbf{1}_{\{\theta_1=\theta^L\}}+c_2^0\mathbf{1}_{\{\theta_1=\theta^H\}}$ as the deployment fee for two types of the defender and $r_0^0:=r_3^0\mathbf{1}_{\{\theta_1=\theta^L\}}+r_4^1\mathbf{1}_{\{\theta_1=\theta^H\}}$ as the penalty for attackers.  
The attacker $\theta_2=\theta^b$ will receive a faked reward $r_5^0>0$ when contacting the avatar, yet he then arrives at an unfavorable state, thus receives limited rewards in the future stages.
The equivalent utility matrix $\tilde{J}_i^0(x^0,a_1^0,a_2^0, \theta_{i},\theta_{-i})$ is shown in Table \ref{table: multiBay-initial}.  
Although the legitimate user can also take action $a_2^0=2$, he should assign zero probability to that action as the payoff is $-\infty$, i.e., a legitimate user should not contact a person that does not exist. 
\begin{table}[h]
\centering
\caption{The utility matrix ($\tilde{J}_1^0,\tilde{J}_2^0$) for player $i=1,2$ under different types.  Although the utility matrix is independent of the current state $x^0$, the action will affect the state transition $f^0$ and then the final state $x^K$ where the utility is state-dependent. 
\label{table: multiBay-initial}}
\begin{tabular}{|c|c|c|c|c|}
\hline
$\theta_2=\theta^g$    & Employee & CEO  & Avatars \\ \hline
NOP          & $(0,r_1^0)$   & $(0, r_1^0)$  & $(0,-\infty)$  \\ \hline
Employee     & $(-c_0^0,r_1^0)$  & $(-c_0^0,r_1^0)$  & $(-c^0_0,-\infty)$\\ \hline
CEO       & $(-c_0^0,r_1^0)$  & $(-c_0^0,r_1^0)$  & $(-c_0^0,-\infty)$ \\ \hline
\end{tabular}

\begin{tabular}{|c|c|c|c|c|}
\hline
$\theta_2=\theta^b$   & Employee & CEO  & Avatars \\ \hline

NOP           & $(-r_2^0,r_2^0)$   & $(-r_2^0, r_2^0)$  & $(0,r_5^0)$  \\ \hline
Employee         & $(-c_0^0,-r_0^0)$  & $(-c_0^0, r_2^0)$  & $(-c^0_0,r_5^0)$\\ \hline
CEO           & $(-c_0^0,r_2^0)$  & $(-c_0^0,-r_0^0)$  & $(-c_1^0,r_5^0)$ \\ \hline
\end{tabular}
\end{table}

Suppose that there are three possible states $\mathcal{X}^1=\{0,1,2\}$ as the output of the initial state transition function $f^0$, i.e., 
user $P_2$ can reach the employee's computer $x^1=1$, the CEO's computer $x^1=2$, or  the honey pot $x^1=0$. 
Assume that the state transition from the initial state $x^0=1$ is determined only by the user's action, i.e., the defender's action does not affect the email delivery from the internal network. On the other hand, the state transition from the external domain  $x^0=0$ is represented as follows. 
If defender chooses not to apply malware analysis system $a_1^0=0$, then user's action $a_2^0=0,1,2$ will lead the initial state $x^0=0$ to state $x^1=1,2,0$, respectively. 
If defender chooses a proactive deployment on the employee's computer $a_1^0=1$, then user's action $a_2^0=0,2$ will drive the initial state $x^0=0$ to state $x^1=0$ and  user's action $a_2^0=1$ will drive the initial state $x^0=0$ to state $x^1=2$. 
The mitigation of the attack is at the tradeoff of blocking some emails from the legitimate user. 
Likewise, if defender chooses a proactive deployment on the CEO's computer $a_1^0=2$, then user's action $a_2^0=1,2$ will lead the initial state $x^0=0$ to state $x^1=0$ and  user's action $a_2^0=0$ will lead the initial state $x^0=0$ to state $x^1=1$. 

\paragraph{Intermediate Stage}
Without loss of generality, we use the privilege escalation scenario in Table \ref{table: muti-stageBay} as the intermediate stage $k=1$. Although the utility matrix is independent of the current state $x^1$, the action will influence the long-term benefit by affecting the state transition $f^1$ as follows. The output state space $\mathcal{X}^K=\{0,1,2,3\}$ represents four different levels of privilege from low to high. 
If the user is at the honeypot $x^1=0$, then he will end up at the honeypot with level-zero privilege $x^K=0$ whatever actions he takes. 
For the user that has arrived at the employee's computer $x^1=1$, if the defender allows privilege escalation $a_1^1=0$, then if the user chooses NOP $a_2^1=0$, the user arrives at level-one privilege $x^K=1$, else if 
 the user requests escalation $a_2^1=1$, he arrives at level-two privilege $x^K=2$.  If the defender restricts the privilege escalation $a_1^1=1$, then $P_2$ arrives at state $x^K=1$ regardless of his action.  
The user arrives at the CEO's computer $x^1=2$ possesses a higher privilege level.  Then, action pair $a_1^1=0, a_2^1=0$ leads to $x^K=2$, and $a_1^1=0, a_2^1=1$ leads to $x^K=3$, and $a_1^1=1, a_2^1=0/1$ leads to $x^K=2$. 

\paragraph{Final Stage}
\begin{table}[h]
\centering
\caption{Two players' utility when the user is either adversarial or legitimate. Define $r^K_0:=r_2^K\mathbf{1}_{\{\theta_1=\theta_1^L\}}+r_3^K\mathbf{1}_{\{\theta_1=\theta_1^H\}}$ as the monitoring reward for two types of systems.}
\begin{tabular}{|l|l|l|l|}
\hline
$\theta_2=\theta^b$& NOP & Access \\ \hline

NOP               & $(0,0)$   & $(r_1^K, r_4^K-r_1^K)$    \\ \hline
Monitor         & $(-c^K,0)$  & $(r^K_0-c^K,-r^K_0)$   \\ \hline
\end{tabular}
\begin{tabular}{|l|l|l|l|}
\hline
$\theta_2=\theta^g$  & NOP & Access \\ \hline
NOP            & $(0,0)$   & $(r_4^K , r_4^K)$    \\ \hline
Monitor        & $(-c^K,0)$  & $(r^K_4-c^K,r^K_4)$   \\ \hline
\end{tabular}
\end{table}

At the final stage $k=K$, we use the Tennessee Eastman (TE) Challenge Process \cite{Ricker} as an example to illustrate how attackers tend to compromise the sensors to cause physical damages (state deviation) of an industrial plant and monetary losses. 
The user's action is to get access to the sensor controller $a_2^K=1$ or not $a_2^K=0$, yet a user at different levels of privilege $x^K$ determines which sensors he can control in the TE process. 
If the attacker changes the sensor reading, the system states such as the pressure and the temperature may deviate from the desired value, which degrades the product quality and even causes the shutdown of the entire process if the deviation exceeds the safety threshold. 
Thus, the shutdown time, as well as the product quality, can be used as the operating reward measure. 
By simulating the TE process, we can determine the reward under the regular operation of the TE process 
$r_4^K(x^K)$ as well as the reward under the compromised sensor readings  $r_1^K(x^K)$. Both $r_4^K$ and $r_1^K$ are a function of the state $x^K$. 
Assume the attacker benefits from the reward reduction under the attacking operation $r_4^K(x^K)-r_1^K(x^K)$ and the system loss under attacks is higher than the monitoring cost $r_4^K(x^K)-r_1^K(x^K)>c^K>0, \forall x^K\in \mathcal{X}^K$. 
On the other hand, the defender chooses to monitor the sensor controller $a_1^K=1$ with a cost $c^K$ or not to monitor $a_1^K=0$. 
Also, we assume $r_3^K>r_2^K>c^K>0$ because the high-type system can collect more information from the monitoring data and the benefit outweighs the monitor cost. 

\section{Conclusion and Future Works}
The area of cybersecurity is an uneven battlefield. 
First, an attacker merely needs to exploit a few vulnerabilities to compromise a system while a defender has to eliminate all potential vulnerabilities. 
Second, the attacker has a plenty of time to study the targeted system yet it is hard for the defender to predict possible settings of attacks until they have happened.
Third, the attacker can be strategic and deceptive and the defender has to adapt to variations and updates of the attacker.
In this chapter, we aim to avoid the route of analyzing every attacks and taking costly  countermeasures. However, we endeavor to tilt the unfavorable situation for the defender by applying a series of game theory models to capture the strategic interactions, the multi-stage persistence, as well as the adversarial and defensive cyber deceptions. 
Future directions include a combination of the theoretical models with data from the simulated or real system under attacks. The analysis of the game theory model provides a theoretic underpinning for our understandings of cybersecurity problems. We can further leverage the scientific and quantitative foundation to investigate mechanism design problems to construct a new battlefield that reverses the attacker's advantage and make the scenario in favor of the defender.

\newpage

\section{Exercise}
\paragraph{QA. Equilibrium Computation and Code Realization.}  
\begin{itemize}
\item[1.] Write a bi-linear program to compute the PBNE of multi-stage game with one-sided incomplete information, i.e., only the user has a type $\theta_1\in \Theta_1$, the defender does not have a type or $P_1$ knows her type. Can you represent it in a matrix form?  (Hint: Corollary $1$ in \cite{huang2018gamesec}.)
\item[2.] Compute the mixed-strategy BNE for the static Bayesian game in Table \ref{table: static Bayesian} with unbiased belief $b_1(\theta^b)=b_1(\theta^g)=0.5$. 
You can program it in Matlab with the toolbox Yalmip\footnote{https://yalmip.github.io/} and a proper nonlinear solver such as Fminicon\footnote{https://www.mathworks.com/help/optim/ug/fmincon.html}.
(Hint: PBNE degenerates to BNE when we take $K=0$.)
\end{itemize}

\paragraph{QB. The Negative Information Gain in Game Theory. }
Let us consider a static Bayesian game with the binary type space $\Theta=\{\theta^1,\theta^2\}$ and initial type belief $b_1(\theta^1)=b_1(\theta^2)=0.5$ as shown in Table \ref{table: exercise}.  Player $1$ is the row player and $P_2$ is the column player. Both players are rational and maximize their own utilities. 

\begin{table}[h]
\caption{A static Bayesian game under two possible types $\theta^1$ and $\theta^2$.
\label{table: exercise} }
\center
\begin{tabular}{ccc}
\hline
$\theta=\theta^1$ & a       & b          \\
A & (10,10) & (18,4)     \\
B & (7,19)  & (17,17)     \\
\hline
\end{tabular}
\quad
\begin{tabular}{ccc}
\hline
$\theta=\theta^2$  & a       & b       \\
A & (10,10) & (18,18) \\
B & (14,18) & (20,20)\\
\hline
\end{tabular}
\end{table}

\begin{itemize}
\item[1.] Compute the BNE strategy and the value of the game, i.e., each player's utility under the BNE strategy. (Hint: you should get a  pure-strategy BNE (B,b) and the value is $(18.5,18.5)$)
\item[2.] Suppose the type value is known to both players, determine the NE under $\theta^1$ and $\theta^2$, respectively. 
\item[3.]  Compute the BNE with one-sided incomplete information, i.e.,  only $P_1$ knows the type value, which is \textit{common knowledge}.  The term \textit{common knowledge} means that $P_1$ knows the type, $P_2$ knows that $P_1$ knows the type, and $P_1$ knows that $P_2$ knows that $P_1$ knows the type, etc. 
\item[4.] Compare the results in question 1-3, does more information always benefit the player with extra information? Can you give an explanation for this negative information gain  in the game setting? 
\end{itemize}

\newpage


\bibliographystyle{spmpsci}      
\bibliography{DeceptionBook}   

\end{document}